\documentclass[aps,prb,10pt,twocolumn,floatfix,superscriptaddress,showpacs,numerical,footinbib]{revtex4-1}
\usepackage[normalem]{ulem}
\usepackage{graphicx}
\usepackage{amsmath}
\usepackage{amssymb}
\usepackage{mathdots}
\usepackage{bbold}
\usepackage[usenames,dvipsnames,pdftex]{color}
\newcommand{\sumnear}{\mathop{\sum}_{\langle i j \rangle}}
\usepackage[%
  colorlinks=true,
  urlcolor=blue,
  linkcolor=blue,
  citecolor=blue
]{hyperref}

\begin{document}
%\begin{CJK*}{GBK}{}
\title{Machine learning of quantum phase transitions}

\author{Xiao-Yu Dong }
\affiliation{Max-Planck-Institut f{\"u}r Physik komplexer Systeme, N{\"o}thnitzer Stra{\ss}e 38, 01187 Dresden, Germany}
\affiliation{Department of Physics and Astronomy, California State University, Northridge, California 91330, USA}
\author{Frank Pollmann}
\affiliation{Technische Universit{\"a}t M{\"u}nchen, Physics Department T42, 85747 Garching, Germany}
\affiliation{Max-Planck-Institut f{\"u}r Physik komplexer Systeme, N{\"o}thnitzer Stra{\ss}e 38, 01187 Dresden, Germany}
\author{Xue-Feng Zhang}
\email{Corresponding authour: zhangxf@cqu.edu.cn}
\affiliation{Department of Physics, Chongqing University, Chongqing 401331, People's Republic of China}
\affiliation{Max-Planck-Institut f{\"u}r Physik komplexer Systeme, N{\"o}thnitzer Stra{\ss}e 38, 01187 Dresden, Germany}

\begin{abstract}
Machine learning algorithms provide a new perspective on the study of physical phenomena.
In this paper, we explore the nature of quantum phase transitions using  multi-color convolutional neural-network (CNN) in
combination with quantum Monte Carlo simulations. 
We propose a method that compresses ($d+1$)-dimensional space-time configurations to a manageable size and then use them as the input for a CNN.
We benchmark our approach on two models and show that both continuous and discontinuous quantum phase transitions can be well detected and characterized. 
Moreover we show that  intermediate phases, which were not trained, can also be identified using our approach. 
\end{abstract}

%\pacs{05.30.Jp, 03.75.Hh, 03.75.Lm, 75.40.Mg, 75.10.Jm}

\maketitle
%\end{CJK*}
% ----------------------------------------------------------------
%\textbf{Introduction.}
%
Machine learning, especially deep learning, has recently shown to be a very powerful tool in the fields of image classification, speech recognition, video activity recognition, machine translation, game playing and so on~\cite{krizhevsky2012imagenet,ML_review,goodfellow2016deep}. 
The basic idea is to train a machine with large datasets such that it can thereafter process and characterize new data. A typical example is image recognition, where a large number of images are used as the training set. During the training process, a non-linear variational function  with images as input and for example the names of objects as output is optimized with respect to a cost function. Using this optimized function, the machine can then recognize the objects in other testing images by knowing their key features.

An important task in condensed matter physics is to characterize different phases of matter and transitions between them~\cite{sachdev2007quantum,wenbook}. Phases can for example be characterized by local order parameters in Landau's theory of spontaneously symmetry breaking~\cite{landau1937theorie}, by topological invariants in topological phases~\cite{zoo_wen}, or by their dynamical properties as in the many-body localized phase \cite{BaskoAleinerAltshuler06}. The main difficulty of this approach is to find characteristic and universal properties of a given phase before we can identify it in a given physical system. In contrast, machine learning techniques promise to classify the phases automatically given a sufficiently large training set is provided.  The deep learning algorithm, which we use in this work, is a method that is capable to learn the key features of individual phases and to classify them directly from ``raw data'' (e.g., the partition function or the ground state wave function). Machine learning is a powerful tool compared to conventional approaches and already inspired physicists to come up with new methods to recognize phases in various settings~\cite{carrasquilla2017machine, PhysRevB.94.195105, PhysRevE.96.022140, PhysRevE.95.062122, PhysRevB.96.184410, PhysRevB.96.144432, wang2018machine, iakovlev2018supervised, kim2018smallest, PhysRevX.7.031038, van2017learning, broecker2017machine, PhysRevB.97.045207, PhysRevB.95.245134, PhysRevB.96.245119, PhysRevLett.118.216401, broecker2017quantum, Rodrigo2018Extrapolating, tanaka2017detection}.

A natural way to use machine learning to identify different phases of matter is with the aid of Monte Carlo method~\cite{mc_binder}. By stochastically moving through configuration space according to a partition function, a large number of samples can be obtained and labelled by different phases. These can then be fed into deep learning algorithms as training sets to classify phases and also detect phase transitions~\cite{carrasquilla2017machine, PhysRevB.94.195105, PhysRevE.96.022140, PhysRevE.95.062122, PhysRevB.96.184410, PhysRevB.96.144432, wang2018machine, iakovlev2018supervised, kim2018smallest}. Considering each sample as a snapshot photo, the classical phase classification is similar to the image recognition.

Quantum Monte Carlo (QMC) methods operate in at least $(d+1)$-dimensional configuration spaces (determinant QMC needs more)~\cite{qmc_review}, where $d$ is the spatial dimension and the extra dimension refers to the imaginary time $\beta = 1/T$ ($T$ is the temperature) direction. To detect quantum phase transitions,  thermal fluctuations have to be strongly suppressed and thus sufficiently low temperatures have to be reached. The resulting huge size of the configuration space, which is proportional to $L^d\beta$ (commonly $\gtrsim$ 100GB~\cite{note}), is too large to be squeezed into a machine learning algorithm. So far only the high temperature regime for small system sizes could be studied by machine learning techniques using the full configuration space~\cite{PhysRevX.7.031038}. Instead, usually different kinds of indirect data are used as input such as the entanglement spectrum~\cite{van2017learning, PhysRevB.95.245134}, Green functions~\cite{broecker2017machine, broecker2017quantum}, winding numbers~\cite{broecker2017quantum} and alike~\cite{PhysRevB.97.045207, PhysRevB.96.245119, PhysRevLett.118.216401}. However, using preprocessed data may cause  important information to be ignored which is contrary to the original idea of machine learning---namely finding characterizing features by itself. Thus it is important to find an efficient machine learning technique that allows to identify quantum phase transitions based on  unfiltered raw data of QMC simulations.

In this paper, we propose a systematic way to compress the $(d+1)$-dimensional configurations such that they can be fed into multi-color conventional neural-networks (CNNs). This approach is inspired by the similarity of the data structure in QMC simulations to video data. Firstly, we investigate the efficiency of our approach by considering the conventional quantum phase transition between Mott-insulator and superfluid. We find that our algorithm correctly identifies it as a continuous quantum phase transition. Secondly, we consider a more complex quantum phase transition where an intermediate supersolid phase emerges. It turns out that our algorithm detects this intermediate phase even though it was not part of the training set. Moreover, we train a CNN to measure winding numbers and it turns out that these can be determined with high accuracy. This indicates that our compression scheme can keep  relevant information in the imaginary time, although the compression ratio (uncompressed/compressed) is more than 300.

\begin{figure}[t]
	\includegraphics[width=0.5\textwidth]{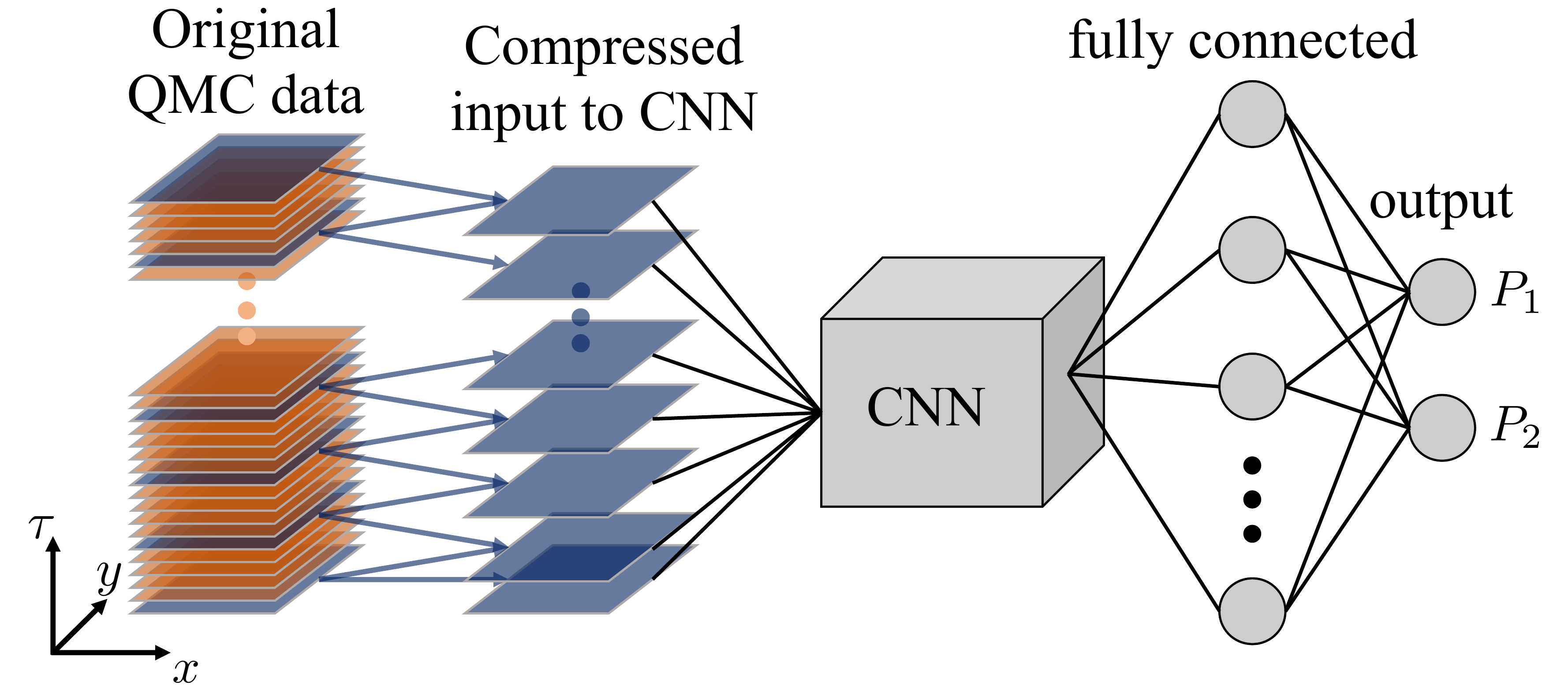}
	\caption{The schematic plot of the data compression process and the structure of the neural networks.
		\label{fig:1}}
\end{figure}

In order to simulate quantum systems, QMC algorithms usually sample the partition function based on an expansion ansatz. For example, in the stochastic series expansion (SEE) for a Hamiltonian of the form $H=\sum_b H_b$ (with $H_b$ are defined on the bonds), the partition function is expanded as 
\begin{eqnarray}
Z &=& \sum_\alpha\sum_{n=0}^{\infty}\sum_{S_n}\frac{(-\beta)^n}{n!}\langle \alpha |\prod_{i=1}^n H_{b_i}|\alpha\rangle,
\label{expansion}
\end{eqnarray}
where $|\alpha\rangle$ is the basis in occupation number representations and $S_n=\{b_1,b_2,...,b_n\}$ is the operator-index sequence~\cite{sse1,sse2,sse3,sse4}. Then we can define a discrete imaginary time $\tau$ such that the state after $\tau$ steps propagation is $|\alpha(\tau)\rangle=\prod_{i=1}^{\tau} H_{b_i}|\alpha\rangle$. Thus, as shown in Fig.~\ref{fig:1}, each term in Eq.~(\ref{expansion}) can be represented as ($d+1$)-dimensional configuration (here $d=2$) in which each slice in imaginary time is $|\alpha(\tau)\rangle$. 

%
%
%--------- figure 1-------------
%--------------------------------
%
%
While diagonal order (e.g., density waves) can be directly detected by the density distribution in each slice $|\alpha(\tau)\rangle$, off-diagonal order (e.g. superfluid order or boson condensation) is characterized by a change between slices. 
Thus the machine learning algorithm should not only consider the information of density distribution at a given time but also the dynamical properties. Each time slice in configuration space can be seen as one frame in a video. To reduce the data size, we can directly borrow the spirit of video compression: The raw data of a video contains a large number of frames. A compressed video keeps only the first frame and  successively the difference between frames. In our problem, we also keep the first time slice $|\alpha(0)\rangle=|\alpha\rangle$. Then, as shown in Fig.~\ref{fig:1}, we divide the whole configuration space into $N$ parts with equal distance in the imaginary time, and only store the difference between the slices at the beginning and end of each part. In this way, the size of input data is manageable. Although it is a lossy compression, we will demonstrate later on that a proper choice of compression strength will keep the relevant information.

After compression, the reduced data will be used to train a deep learning model for classification. The structure we used is demonstrated in Fig.~\ref{fig:1}. 
The model has $N+1$ input channels, which are called color channels. The $N+1$ matrices obtained by compressing the QMC data are fed into these channels one-by-one in sequence. Then, the data in all channels is loaded into a CNN. Two standard CNN layers are used in this work. After that, two fully connected layers are followed before it is fed into the final output layer with two neurons. The number of output neurons here is determined by the number $m$ of different phase labels of the input data. Their values, denoted as $P_i$ with $i=1,2,...,m$, correspond to the probability of $i$-th quantum phase, respectively, and $\sum_{i=1}^m P_i=1$. In this work, we use two classes of data as input and thus $m=2$.
%
%--------- figure 2-------------
\begin{figure}[t]
	\includegraphics[width=\columnwidth]{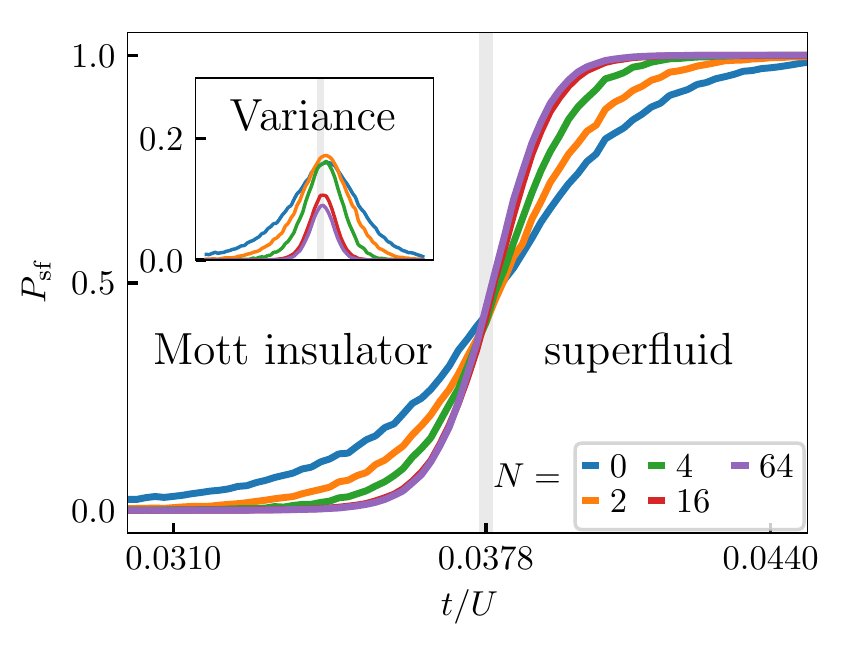}
	\caption{Probabilities of the superfluid phase $P_{\mathrm{sf}}$ for the  Bose-Hubbard model on the triangular lattice with different values of $N$ and $t/U$ at $\beta=100$ and $L=12$. Inset: the corresponding variances of the predictions.
		\label{fig:2}}
\end{figure}
%--------------------------------
%

%\textbf{Mott insulator-superfluid phase transition.} 
%
Firstly, we consider a continuous quantum phase transition between a Mott-insulator and superfluid phase of the Bose-Hubbard model on the triangular lattice. The Hamiltonian reads
\begin{eqnarray}
H &=& -t \sum_{\langle i, j\rangle}(b_{i}^{\dag}b_{j}+h.c.) + \frac U 2 \mathop{\sum}_in_i(n_i-1) 
 \label{BH}
\end{eqnarray}
where $b_{i}^{\dag}$($b_{i}$) is the creation (annihilation) operator a boson, $\langle i,j\rangle$ represent nearest-neighbour sites, $t$ is the hopping strength, and $U$ describes the on-site repulsion. This model is relevant in the context of ultra-cold atoms on a triangular optical lattices~\cite{sengstock}. At small $|t/U|$ and commensurate filling, the system is in a Mott-insulating phase with finite energy gap. When sufficiently increasing $|t/U|$, quantum fluctuations yield a phase transition into a gapless superfluid phase. The phase transition belongs to a 3D XY type universality class~\cite{fisher}, and the critical point at filling $\langle n\rangle=1$ is approximately $t_c/U\approx0.0378$~\cite{AF_tri}.

Using QMC, we produce 20000 samples for each $t/U$ in the region $t/U=[0.0300,0.0450]$ with step $\delta(t/U)=0.0002$, the length of imaginary time is around $5\times 10^4$. After compressing these samples with different $N$ (which reflect the compression strength), we separate the samples at each point into two sets. One set is used for training, and the other  is used for testing. The samples in the training data set are collected  deep within the two phases: $t/U=[0.0300, 0.0330]$ for the Mott-insulator phase and $[0.0420, 0.0450]$ for the superfluid phase. While the testing data set contains the samples from the whole region including the phase transition point. When the prediction error on the training set is converged, the testing data set is used to produce predictions in the whole region. The probabilities of the superfluid phase $P_{\textrm{sf}}$  are shown in Fig.~\ref{fig:2} for different $N$ and $t/U$. The steepness of the curve near the transition is an indicator for the quality of the prediction. We can see that the lines become steeper with increasing $N$ of the input data, and also the lines in each region are closer to the perfect prediction values $0$ or $1$. Moreover, if we check the variances of the predictions, it also becomes smaller with increasing $N$ as shown in the inset of Fig.~\ref{fig:2}. This means that enlarging $N$ will improve the accuracy of the predictions---this is indeed expected as a larger $N$ implies less compression and more accurate data to process.  Note that the Mott insulating and superfluid phases are distinguished by the superfluid density, which can be directly related to the winding of configuration space in the imaginary time direction. Extracting this winding requires knowledge about the imaginary time direction and cannot be extracted from a single time slice. Moreover, we  find that all the curves approximately cross at the critical point, and $N=64$ is nearly converged.

%--------- figure 3-------------
\begin{figure}[t]
	\includegraphics[width=\columnwidth]{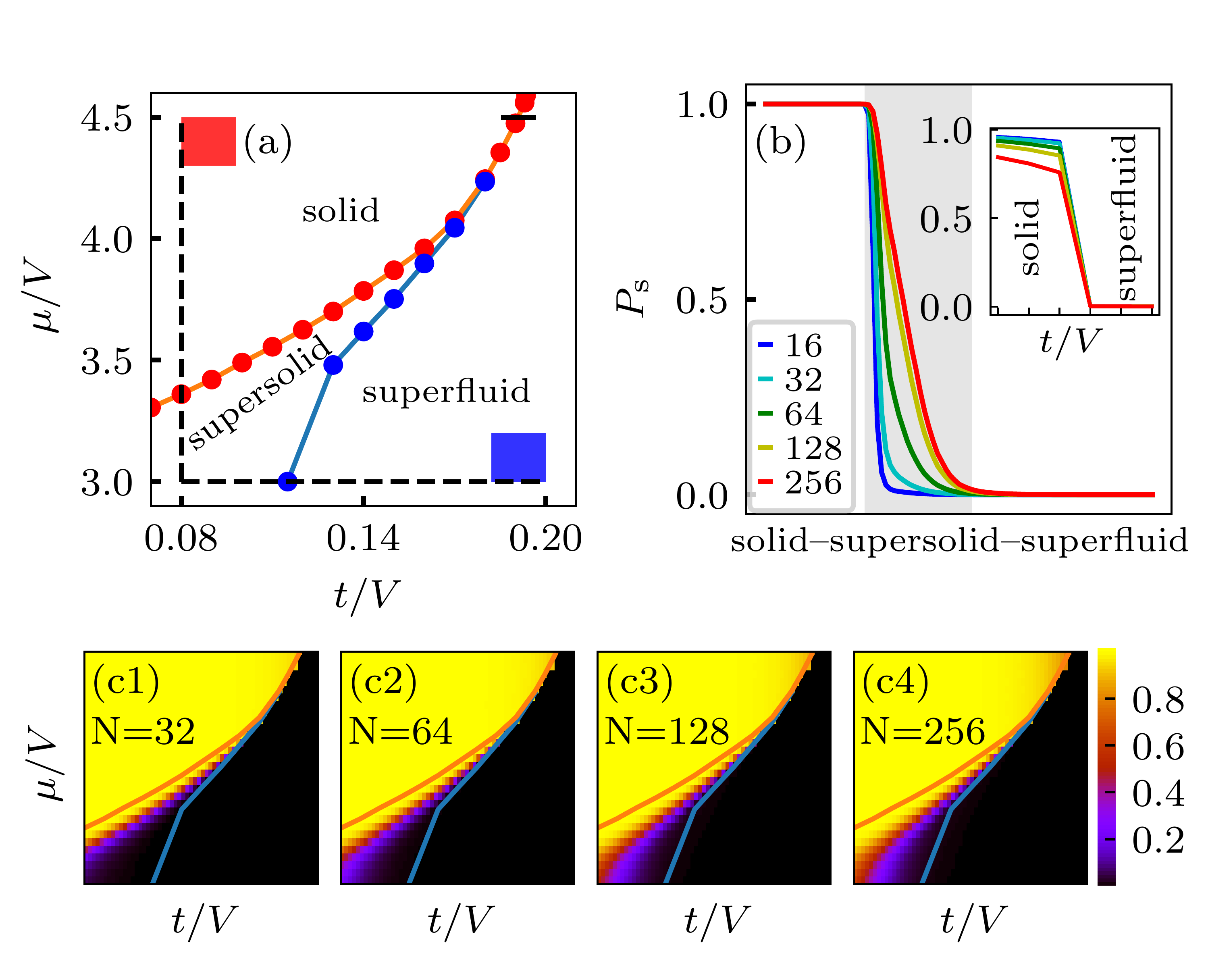}
	\caption{(a) Phase diagram of the extended hard-core Bose-Hubbard model on the triangular lattice (with size $L\times L, L=12$) calculated by QMC \cite{ss10}. The red and blue rectangles are the regions which we use as training set. (b) The predictions of the probabilities of the solid phase with different $N$ on the trajectory along the black dashed line in (a). The inset is the predictions of the probabilities of the solid phase during the first order phase transition along the trajectory marked with the black line ($\mu/V=4.5$ and $t/V\in [0.186, 0.196]$ with step $\delta(t/V) = 0.002$) at the right top corner in (a). (c1-c4) Predicted phase diagram with $N= 32$, $64$, $128$, and $256$, respectively. 
		\label{fig:3}}
\end{figure}
%--------------------------------

%\textbf{Signatures of the supersolid phase.}
%
The above example shows that the deep learning model can well predict the quantum phases and related continuous phase transition. However, the quantum phase transition could also be first order or an intermediate phase could emerge between two phases. A question we address now is whether deep learning can predict the existence of the intermediate phase without ``knowing" it. For this we consider an extended hard-core Bose-Hubbard model on the triangular lattice:
\begin{eqnarray}
H = -t \sumnear(b_{i}^{\dag}b_{j}+h.c.) + V \sumnear n_in_j + \mu\sum_i n_i
\label{eBH}
\end{eqnarray}
where $V$ denotes the repulsive interaction between nearest-neighbor sites, hard-core implies that only occupancies $n_i=0,1$ are allowed, and $\mu$ is the chemical potential. The phase diagram is shown in Fig.~\ref{fig:3}(a). The solid phase breaks the translational symmetry and the superfluid phase breaks the U(1) symmetry. Interesting, there is an intermediate \emph{supersolid} phase which breaks both symmetries~\cite{ss1,ss2,ss3,ss4,ss5,ss6,ss7,ss8,ss9,ss10}. The triangular lattice is composed of three sublattices, the solid phase can be viewed as two sublattices being fully occupied and the other one is empty. A qualitative picture of the supersolid phase is a doped solid with holes that can move on a honeycomb sublattice~\cite{ss10}. 

Following the same strategy as above, we collect the samples deep in phases from regions marked with color blocks in Fig.~\ref{fig:3}(a), and the length of imaginary time is around $7\times 10^4$. After data compression, we feed them into the deep learning model for training. Next we run the prediction in the whole parameter region. As shown in the inset of Fig.~\ref{fig:3}(b), a first order phase transition is clearly reflected by the sudden jump of the prediction of the probabilities of the solid phase. Because of the large difference between solid and superfluid phase, even small $N$ can present the discontinuity of the phase transition. In order to check the behavior of the deep learning model to the ``unknown" supersolid phase, we choose the ``L" shape trajectory, going from $\mu/U = 4.5$ to $3.0$ with fixed $t/U = 0.08$, and then going from $t/U = 0.08$ to $0.20$ with fixed $\mu/U = 3.0$. The predictions of the probabilities of the solid phase $P_{\mathrm{s}}$ is plotted in Fig.~\ref{fig:3}(b) (the probability of superfluid phase is $P_{\mathrm{sf}} =1-P_{\mathrm{s}}$). In contrast to the continuous quantum phase transition, decreasing the compression rate makes the curve more smooth. In other words, the deep learning model becomes more ``confused" in the supersolid phase when taking into account more data. If we label region with $P_{\mathrm{s}}\in (0,1)$ as the intermediate region, from the predictions on the whole phase diagram in Fig.~\ref{fig:3}(c1-c4) with $N = 32, 64, 128, 256$, we can find the intermediate region approaches to the real boundary of the supersolid phase. The neural network has been trained to  recognize the superfluid and solid phase. When it faces the co-existing order in the intermediate phase, the different contributions from the supersolid will intensify corresponding output neurons such that the prediction value is neither zero nor one. As complementary test, we  performed simulations with three output neurons--including one additional neuron to learn the intermediate phase. Using an enlarged training set, we find a phase diagram identical to the one identified by the ``confusion'' approach discussed above (see supplemental material for details).

In brief, the relation between prediction of probability and compression strength can be used to distinguish direct quantum phase transition and  intermediate phases. The reason such compression can keep the key information in imaginary time is its equivalence to the high frequency truncation. When we increase the value of $N$, more details of short distance in the imaginary time will be captured, implying that higher frequency information will be included. Considering that low frequencies is usually more relevant at low energies, such a compression method has substantial advantages for studying the quantum phase transition. 

At last, we try to directly check to which extend the winding in imaginary time can be extracted. In QMC, the bosons can form a net current flowing around the periodic system in real space. Due to the periodic boundary condition in imaginary time (trace of partition function), such current  can only wind around the system an integer number of times in real direction. This integer number is called  the ``winding number''. This  quantity represents  non-local properties of the quantum system and is proportional to the superfluid density~\cite{rhos}.

%--------- figure 4-------------
\begin{figure}[h]
\includegraphics[width=\columnwidth]{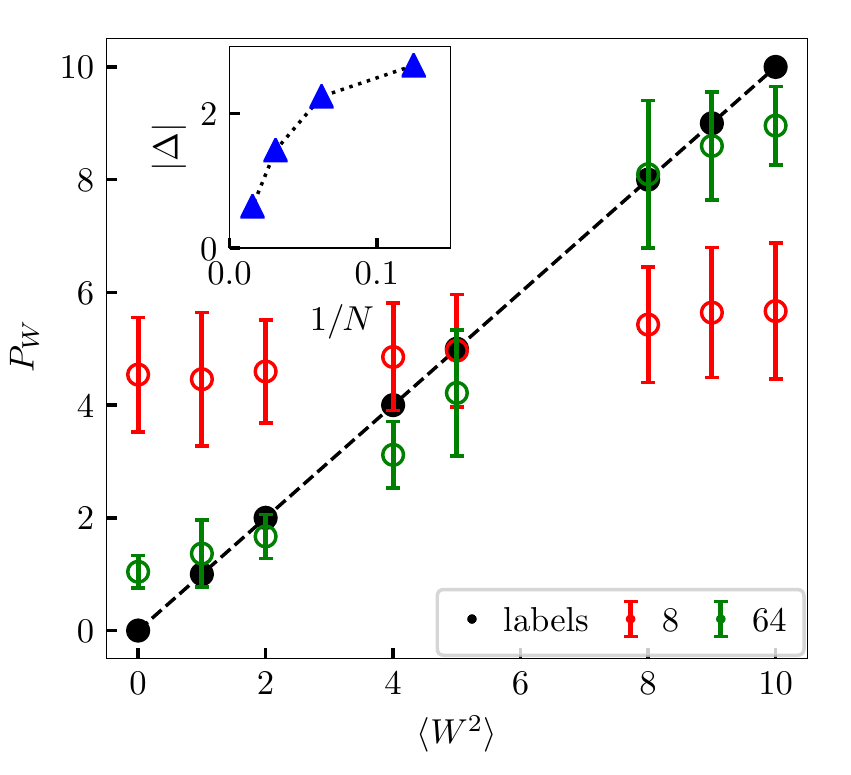}
\caption{The predictions of the winding numbers $W^2$ by regression method with $N=8$ and $64$. The black dots with dashed lines are the exact numbers of $W^2$. The errorbars denote the standard deviations of the predictions on the test set. Inset: The mean of the absolute difference between the labels and the predictions versus $1/N$.
\label{fig:4}}
\end{figure}
%--------------------------------

In order to count the winding number using deep learning techinques, the neural network with one output neuron is used. We then randomly select the samples with small winding numbers from the samples of the previously discussed Bose-Hubbard model and divide them into a new training and testing set. Meanwhile, the labels are changed into the square sum of the winding numbers in $x$- and $y$-direction $W^2 = W_x^2 + W_y^2$. After training with a regression algorithm, the output values of the testing set are the predicted winding numbers $W^2$. The results with $N = 8$ and $64$ are shown in Fig.~\ref{fig:4} and the black dots with dashed line denotes the exact values of $W^2$. Clearly, the predictions with less compression, i.e. $N=64$, are closer to the exact values compared to the high compression with $N = 8$. In addition, we also plot the average absolute difference $|\Delta|$ between the predictions and exact values versus $1/N$ in the inset of Fig.~\ref{fig:4}, and it tends to zero when decreasing the compression strength. The winding number prediction gives another strong verification that the data compression and deep learning model can not only catch the long range correlation, but also even the winding in imaginary time.

In conclusion, we proposed a systematic way of generating and compressing training samples to be used for machine learning in combination with quantum Monte Carlo methods. The neural networks for deep learning are composed of multi-color CNN following with fully connected neuron layers. By implementing this method for two types of Bose-Hubbard models on the triangular lattice, we found a  qualitatively distinct behavior for the different cases: (1) for  a first order phase transition, the machine learning model can well predict the quantum phases even for strong compression, (2) for the continuous case, the prediction of the probability of one phase becomes steeper when decreasing the compression strength; (3) if there exists an intermediate phase, the prediction shows the opposite behavior, i.e., the slope becomes more gradual when decreasing the compression strength. We also tested the winding number predictions with a regression algorithms, and its high accuracy suggests that our method obtains relevant information about the topological properties in  the imaginary time direction. We argue that such a deep learning method  recognizes quantum phase transitions well because the compression scheme only removes the high frequency part. Our ab-initio approach for deep learning of quantum phase transitions could also be extended to other world-line based quantum Monte Carlo simulation, and will be helpful for detecting unknown phases where a proper order parameter is unknown (e.g., spin liquids or many-body localization). Moreover, it may shed light on the ``prediction" of supervised machine learning.

Related work: While completing this manuscript, we became aware of a related work~\cite{hsu2018machine} which shows that supervised machine learning can be used to detect novel phases that have not been trained in the context of many-body localization.

\section*{Acknowledgements} 
We thanks Chen Zhang, Zhi-Yuan Xie for helpful discussions, and Hubert Scherrer-Paulus for technical supports on Google Tensorflow and GPU. F. Pollmann acknowledges support from DFG through Research Unit FOR 1807 with grant no.\ PO 1370/2-1 and from the Nanosystems Initiative Munich (NIM) by the German Excellence Inititiative, the European Research Council (ERC) under the European Union's Horizon 2020 research and innovation programme (grant agreement no. 771537). X.-F. Zhang acknowledges funding from Project No. 2018CDQYWL0047 supported by the Fundamental Research Funds for the Central Universities, Grant No. cstc2018jcyjAX0399 by Chongqing Natural Science Foundation and from the National Science Foundation of China under Grants No. 11804034 and No. 11874094.

%\appendix
\begin{appendix}
\section{Prediction of the intermediate supersolid phase}
%--------- figure S1-------------
\begin{figure}[h]
	\includegraphics[width=0.45\textwidth]{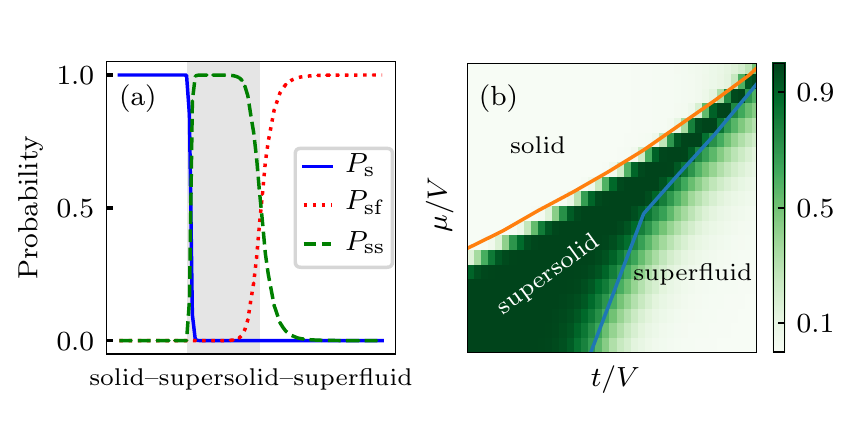}
	\caption{(a) The probabilities of the solid (blue solid line), superfluid (red dotted line), and supersolid (green dashed line) phases on the trajectory along the black dashed line in Fig.~3(a) of the main text, with $N=64$. (b) The phase diagram with colors denoting the probability of the supersolid phase $P_{\mathrm{ss}}$.}
		\label{fig:S1}
\end{figure}
%--------------------------------

In this supplementary material, we explain the details about the calculations which give the probability of the supersolid phase of the extended hard-core Bose-Hubbard model on the triangular lattice. From the results of the main text, we get the signature of the intermediate supersolid phase by inspecting the dependence of predictions on the compression ratio of the input data. The structure of the neural network is fixed when we train it, so if the training set has only two classes of data from two phases, the two output neurons could only give the probabilities of these two phases. Therefore, it is almost impossible to predict the existence of a third class which has no samples in the training set. For example, a neural network could not recognize a picture of a monkey, if it is trained by only the pictures of cats and dogs. Thanks to the tuning parameter $N$ in our proposal, the existence of the third phase could be observed. However, the direct probability of the third phase could not be given in such a fixed structure. To prove the existence of the intermediate supersolid phase, we could collect the data for this phase in the region we inspected before (the data in region $t/V\in[0.080,0.098]$ and $\mu/V\in[3.00,3.20]$ are used), and add them to the training data, too. A third label is given to them to distinguish from the other two phases. Now the number of the output neurons is three, and the probability of the supersolid phase $P_{\mathrm{ss}}$ could be obtained directly. In Fig.~\ref{fig:S1}(a), the predictions of these three phases on the trajectory along the black dashed line in Fig.~3(a) of the main text are given. It shows clearly that there are three phases separated by two phase transition points. The region of the supersolid phase is given in Fig.~\ref{fig:S1}(b) with the color denoting the probability of the supersolid phase.
\end{appendix}
\bibliography{ML}

\begin{thebibliography}{49}%
\makeatletter
\providecommand \@ifxundefined [1]{%
 \@ifx{#1\undefined}
}%
\providecommand \@ifnum [1]{%
 \ifnum #1\expandafter \@firstoftwo
 \else \expandafter \@secondoftwo
 \fi
}%
\providecommand \@ifx [1]{%
 \ifx #1\expandafter \@firstoftwo
 \else \expandafter \@secondoftwo
 \fi
}%
\providecommand \natexlab [1]{#1}%
\providecommand \enquote  [1]{``#1''}%
\providecommand \bibnamefont  [1]{#1}%
\providecommand \bibfnamefont [1]{#1}%
\providecommand \citenamefont [1]{#1}%
\providecommand \href@noop [0]{\@secondoftwo}%
\providecommand \href [0]{\begingroup \@sanitize@url \@href}%
\providecommand \@href[1]{\@@startlink{#1}\@@href}%
\providecommand \@@href[1]{\endgroup#1\@@endlink}%
\providecommand \@sanitize@url [0]{\catcode `\\12\catcode `\$12\catcode
  `\&12\catcode `\#12\catcode `\^12\catcode `\_12\catcode `\%12\relax}%
\providecommand \@@startlink[1]{}%
\providecommand \@@endlink[0]{}%
\providecommand \url  [0]{\begingroup\@sanitize@url \@url }%
\providecommand \@url [1]{\endgroup\@href {#1}{\urlprefix }}%
\providecommand \urlprefix  [0]{URL }%
\providecommand \Eprint [0]{\href }%
\providecommand \doibase [0]{http://dx.doi.org/}%
\providecommand \selectlanguage [0]{\@gobble}%
\providecommand \bibinfo  [0]{\@secondoftwo}%
\providecommand \bibfield  [0]{\@secondoftwo}%
\providecommand \translation [1]{[#1]}%
\providecommand \BibitemOpen [0]{}%
\providecommand \bibitemStop [0]{}%
\providecommand \bibitemNoStop [0]{.\EOS\space}%
\providecommand \EOS [0]{\spacefactor3000\relax}%
\providecommand \BibitemShut  [1]{\csname bibitem#1\endcsname}%
\let\auto@bib@innerbib\@empty
%</preamble>
\bibitem [{\citenamefont {Krizhevsky}\ \emph {et~al.}(2012)\citenamefont
  {Krizhevsky}, \citenamefont {Sutskever},\ and\ \citenamefont
  {Hinton}}]{krizhevsky2012imagenet}%
  \BibitemOpen
  \bibfield  {author} {\bibinfo {author} {\bibfnamefont {A.}~\bibnamefont
  {Krizhevsky}}, \bibinfo {author} {\bibfnamefont {I.}~\bibnamefont
  {Sutskever}}, \ and\ \bibinfo {author} {\bibfnamefont {G.~E.}\ \bibnamefont
  {Hinton}},\ }in\ \href@noop {} {\emph {\bibinfo {booktitle} {Advances in
  neural information processing systems}}}\ (\bibinfo {year} {2012})\ pp.\
  \bibinfo {pages} {1097--1105}\BibitemShut {NoStop}%
\bibitem [{\citenamefont {LeCun}\ \emph {et~al.}(2015)\citenamefont {LeCun},
  \citenamefont {Bengio},\ and\ \citenamefont {Hinton}}]{ML_review}%
  \BibitemOpen
  \bibfield  {author} {\bibinfo {author} {\bibfnamefont {Y.}~\bibnamefont
  {LeCun}}, \bibinfo {author} {\bibfnamefont {Y.}~\bibnamefont {Bengio}}, \
  and\ \bibinfo {author} {\bibfnamefont {G.}~\bibnamefont {Hinton}},\ }\href
  {\doibase 10.1038/nature14539} {\bibfield  {journal} {\bibinfo  {journal}
  {Nature}\ }\textbf {\bibinfo {volume} {521}},\ \bibinfo {pages} {436}
  (\bibinfo {year} {2015})}\BibitemShut {NoStop}%
\bibitem [{\citenamefont {Goodfellow}\ \emph {et~al.}(2016)\citenamefont
  {Goodfellow}, \citenamefont {Bengio}, \citenamefont {Courville},\ and\
  \citenamefont {Bengio}}]{goodfellow2016deep}%
  \BibitemOpen
  \bibfield  {author} {\bibinfo {author} {\bibfnamefont {I.}~\bibnamefont
  {Goodfellow}}, \bibinfo {author} {\bibfnamefont {Y.}~\bibnamefont {Bengio}},
  \bibinfo {author} {\bibfnamefont {A.}~\bibnamefont {Courville}}, \ and\
  \bibinfo {author} {\bibfnamefont {Y.}~\bibnamefont {Bengio}},\ }\href@noop {}
  {\emph {\bibinfo {title} {Deep learning}}},\ Vol.~\bibinfo {volume} {1}\
  (\bibinfo  {publisher} {MIT press Cambridge},\ \bibinfo {year}
  {2016})\BibitemShut {NoStop}%
\bibitem [{\citenamefont {Sachdev}(2007)}]{sachdev2007quantum}%
  \BibitemOpen
  \bibfield  {author} {\bibinfo {author} {\bibfnamefont {S.}~\bibnamefont
  {Sachdev}},\ }\href@noop {} {\emph {\bibinfo {title} {Quantum phase
  transitions}}}\ (\bibinfo  {publisher} {Wiley Online Library},\ \bibinfo
  {year} {2007})\BibitemShut {NoStop}%
\bibitem [{\citenamefont {Wen}(2004)}]{wenbook}%
  \BibitemOpen
  \bibfield  {author} {\bibinfo {author} {\bibfnamefont {X.-G.}\ \bibnamefont
  {Wen}},\ }\href
  {http://libproxy.mit.edu/login?url=http://search.ebscohost.com/login.aspx?direct=true&db=nlebk&AN=186592&site=ehost-live}
  {\emph {\bibinfo {title} {Quantum Field Theory of Many-body Systems : From
  the Origin of Sound to an Origin of Light and Electrons.}}},\ Oxford Graduate
  Texts\ (\bibinfo  {publisher} {OUP Premium},\ \bibinfo {year}
  {2004})\BibitemShut {NoStop}%
\bibitem [{\citenamefont {Landau}(1937)}]{landau1937theorie}%
  \BibitemOpen
  \bibfield  {author} {\bibinfo {author} {\bibfnamefont {L.}~\bibnamefont
  {Landau}},\ }\href@noop {} {\bibfield  {journal} {\bibinfo  {journal} {Phys.
  Z. Sowjetunion}\ }\textbf {\bibinfo {volume} {11}},\ \bibinfo {pages} {26}
  (\bibinfo {year} {1937})}\BibitemShut {NoStop}%
\bibitem [{\citenamefont {Wen}(2017)}]{zoo_wen}%
  \BibitemOpen
  \bibfield  {author} {\bibinfo {author} {\bibfnamefont {X.-G.}\ \bibnamefont
  {Wen}},\ }\href {\doibase 10.1103/RevModPhys.89.041004} {\bibfield  {journal}
  {\bibinfo  {journal} {Rev. Mod. Phys.}\ }\textbf {\bibinfo {volume} {89}},\
  \bibinfo {pages} {041004} (\bibinfo {year} {2017})}\BibitemShut {NoStop}%
\bibitem [{\citenamefont {Basko}\ \emph {et~al.}(2006)\citenamefont {Basko},
  \citenamefont {Aleiner},\ and\ \citenamefont
  {Altshuler}}]{BaskoAleinerAltshuler06}%
  \BibitemOpen
  \bibfield  {author} {\bibinfo {author} {\bibfnamefont {D.}~\bibnamefont
  {Basko}}, \bibinfo {author} {\bibfnamefont {I.}~\bibnamefont {Aleiner}}, \
  and\ \bibinfo {author} {\bibfnamefont {B.}~\bibnamefont {Altshuler}},\ }\href
  {\doibase http://dx.doi.org/10.1016/j.aop.2005.11.014} {\bibfield  {journal}
  {\bibinfo  {journal} {Annals of Physics}\ }\textbf {\bibinfo {volume}
  {321}},\ \bibinfo {pages} {1126 } (\bibinfo {year} {2006})}\BibitemShut
  {NoStop}%
\bibitem [{\citenamefont {Carrasquilla}\ and\ \citenamefont
  {Melko}(2017)}]{carrasquilla2017machine}%
  \BibitemOpen
  \bibfield  {author} {\bibinfo {author} {\bibfnamefont {J.}~\bibnamefont
  {Carrasquilla}}\ and\ \bibinfo {author} {\bibfnamefont {R.~G.}\ \bibnamefont
  {Melko}},\ }\href@noop {} {\bibfield  {journal} {\bibinfo  {journal} {Nature
  Physics}\ }\textbf {\bibinfo {volume} {13}},\ \bibinfo {pages} {431}
  (\bibinfo {year} {2017})}\BibitemShut {NoStop}%
\bibitem [{\citenamefont {Wang}(2016)}]{PhysRevB.94.195105}%
  \BibitemOpen
  \bibfield  {author} {\bibinfo {author} {\bibfnamefont {L.}~\bibnamefont
  {Wang}},\ }\href {\doibase 10.1103/PhysRevB.94.195105} {\bibfield  {journal}
  {\bibinfo  {journal} {Phys. Rev. B}\ }\textbf {\bibinfo {volume} {94}},\
  \bibinfo {pages} {195105} (\bibinfo {year} {2016})}\BibitemShut {NoStop}%
\bibitem [{\citenamefont {Wetzel}(2017)}]{PhysRevE.96.022140}%
  \BibitemOpen
  \bibfield  {author} {\bibinfo {author} {\bibfnamefont {S.~J.}\ \bibnamefont
  {Wetzel}},\ }\href {\doibase 10.1103/PhysRevE.96.022140} {\bibfield
  {journal} {\bibinfo  {journal} {Phys. Rev. E}\ }\textbf {\bibinfo {volume}
  {96}},\ \bibinfo {pages} {022140} (\bibinfo {year} {2017})}\BibitemShut
  {NoStop}%
\bibitem [{\citenamefont {Hu}\ \emph {et~al.}(2017)\citenamefont {Hu},
  \citenamefont {Singh},\ and\ \citenamefont {Scalettar}}]{PhysRevE.95.062122}%
  \BibitemOpen
  \bibfield  {author} {\bibinfo {author} {\bibfnamefont {W.}~\bibnamefont
  {Hu}}, \bibinfo {author} {\bibfnamefont {R.~R.~P.}\ \bibnamefont {Singh}}, \
  and\ \bibinfo {author} {\bibfnamefont {R.~T.}\ \bibnamefont {Scalettar}},\
  }\href {\doibase 10.1103/PhysRevE.95.062122} {\bibfield  {journal} {\bibinfo
  {journal} {Phys. Rev. E}\ }\textbf {\bibinfo {volume} {95}},\ \bibinfo
  {pages} {062122} (\bibinfo {year} {2017})}\BibitemShut {NoStop}%
\bibitem [{\citenamefont {Wetzel}\ and\ \citenamefont
  {Scherzer}(2017)}]{PhysRevB.96.184410}%
  \BibitemOpen
  \bibfield  {author} {\bibinfo {author} {\bibfnamefont {S.~J.}\ \bibnamefont
  {Wetzel}}\ and\ \bibinfo {author} {\bibfnamefont {M.}~\bibnamefont
  {Scherzer}},\ }\href {\doibase 10.1103/PhysRevB.96.184410} {\bibfield
  {journal} {\bibinfo  {journal} {Phys. Rev. B}\ }\textbf {\bibinfo {volume}
  {96}},\ \bibinfo {pages} {184410} (\bibinfo {year} {2017})}\BibitemShut
  {NoStop}%
\bibitem [{\citenamefont {Wang}\ and\ \citenamefont
  {Zhai}(2017)}]{PhysRevB.96.144432}%
  \BibitemOpen
  \bibfield  {author} {\bibinfo {author} {\bibfnamefont {C.}~\bibnamefont
  {Wang}}\ and\ \bibinfo {author} {\bibfnamefont {H.}~\bibnamefont {Zhai}},\
  }\href {\doibase 10.1103/PhysRevB.96.144432} {\bibfield  {journal} {\bibinfo
  {journal} {Phys. Rev. B}\ }\textbf {\bibinfo {volume} {96}},\ \bibinfo
  {pages} {144432} (\bibinfo {year} {2017})}\BibitemShut {NoStop}%
\bibitem [{\citenamefont {Wang}\ and\ \citenamefont
  {Zhai}(2018)}]{wang2018machine}%
  \BibitemOpen
  \bibfield  {author} {\bibinfo {author} {\bibfnamefont {C.}~\bibnamefont
  {Wang}}\ and\ \bibinfo {author} {\bibfnamefont {H.}~\bibnamefont {Zhai}},\
  }\href@noop {} {\bibfield  {journal} {\bibinfo  {journal} {arXiv preprint
  arXiv:1803.01205}\ } (\bibinfo {year} {2018})}\BibitemShut {NoStop}%
\bibitem [{\citenamefont {Iakovlev}\ \emph {et~al.}(2018)\citenamefont
  {Iakovlev}, \citenamefont {Sotnikov},\ and\ \citenamefont
  {Mazurenko}}]{iakovlev2018supervised}%
  \BibitemOpen
  \bibfield  {author} {\bibinfo {author} {\bibfnamefont {I.}~\bibnamefont
  {Iakovlev}}, \bibinfo {author} {\bibfnamefont {O.}~\bibnamefont {Sotnikov}},
  \ and\ \bibinfo {author} {\bibfnamefont {V.}~\bibnamefont {Mazurenko}},\
  }\href@noop {} {\bibfield  {journal} {\bibinfo  {journal} {arXiv preprint
  arXiv:1803.06682}\ } (\bibinfo {year} {2018})}\BibitemShut {NoStop}%
\bibitem [{\citenamefont {Kim}\ and\ \citenamefont
  {Kim}(2018)}]{kim2018smallest}%
  \BibitemOpen
  \bibfield  {author} {\bibinfo {author} {\bibfnamefont {D.}~\bibnamefont
  {Kim}}\ and\ \bibinfo {author} {\bibfnamefont {D.-H.}\ \bibnamefont {Kim}},\
  }\href@noop {} {\bibfield  {journal} {\bibinfo  {journal} {arXiv preprint
  arXiv:1804.02171}\ } (\bibinfo {year} {2018})}\BibitemShut {NoStop}%
\bibitem [{\citenamefont {Ch'ng}\ \emph {et~al.}(2017)\citenamefont {Ch'ng},
  \citenamefont {Carrasquilla}, \citenamefont {Melko},\ and\ \citenamefont
  {Khatami}}]{PhysRevX.7.031038}%
  \BibitemOpen
  \bibfield  {author} {\bibinfo {author} {\bibfnamefont {K.}~\bibnamefont
  {Ch'ng}}, \bibinfo {author} {\bibfnamefont {J.}~\bibnamefont {Carrasquilla}},
  \bibinfo {author} {\bibfnamefont {R.~G.}\ \bibnamefont {Melko}}, \ and\
  \bibinfo {author} {\bibfnamefont {E.}~\bibnamefont {Khatami}},\ }\href
  {\doibase 10.1103/PhysRevX.7.031038} {\bibfield  {journal} {\bibinfo
  {journal} {Phys. Rev. X}\ }\textbf {\bibinfo {volume} {7}},\ \bibinfo {pages}
  {031038} (\bibinfo {year} {2017})}\BibitemShut {NoStop}%
\bibitem [{\citenamefont {van Nieuwenburg}\ \emph {et~al.}(2017)\citenamefont
  {van Nieuwenburg}, \citenamefont {Liu},\ and\ \citenamefont
  {Huber}}]{van2017learning}%
  \BibitemOpen
  \bibfield  {author} {\bibinfo {author} {\bibfnamefont {E.~P.}\ \bibnamefont
  {van Nieuwenburg}}, \bibinfo {author} {\bibfnamefont {Y.-H.}\ \bibnamefont
  {Liu}}, \ and\ \bibinfo {author} {\bibfnamefont {S.~D.}\ \bibnamefont
  {Huber}},\ }\href@noop {} {\bibfield  {journal} {\bibinfo  {journal} {Nature
  Physics}\ }\textbf {\bibinfo {volume} {13}},\ \bibinfo {pages} {435}
  (\bibinfo {year} {2017})}\BibitemShut {NoStop}%
\bibitem [{\citenamefont {Broecker}\ \emph
  {et~al.}(2017{\natexlab{a}})\citenamefont {Broecker}, \citenamefont
  {Carrasquilla}, \citenamefont {Melko},\ and\ \citenamefont
  {Trebst}}]{broecker2017machine}%
  \BibitemOpen
  \bibfield  {author} {\bibinfo {author} {\bibfnamefont {P.}~\bibnamefont
  {Broecker}}, \bibinfo {author} {\bibfnamefont {J.}~\bibnamefont
  {Carrasquilla}}, \bibinfo {author} {\bibfnamefont {R.~G.}\ \bibnamefont
  {Melko}}, \ and\ \bibinfo {author} {\bibfnamefont {S.}~\bibnamefont
  {Trebst}},\ }\href@noop {} {\bibfield  {journal} {\bibinfo  {journal}
  {Scientific reports}\ }\textbf {\bibinfo {volume} {7}},\ \bibinfo {pages}
  {8823} (\bibinfo {year} {2017}{\natexlab{a}})}\BibitemShut {NoStop}%
\bibitem [{\citenamefont {Beach}\ \emph {et~al.}(2018)\citenamefont {Beach},
  \citenamefont {Golubeva},\ and\ \citenamefont {Melko}}]{PhysRevB.97.045207}%
  \BibitemOpen
  \bibfield  {author} {\bibinfo {author} {\bibfnamefont {M.~J.~S.}\
  \bibnamefont {Beach}}, \bibinfo {author} {\bibfnamefont {A.}~\bibnamefont
  {Golubeva}}, \ and\ \bibinfo {author} {\bibfnamefont {R.~G.}\ \bibnamefont
  {Melko}},\ }\href {\doibase 10.1103/PhysRevB.97.045207} {\bibfield  {journal}
  {\bibinfo  {journal} {Phys. Rev. B}\ }\textbf {\bibinfo {volume} {97}},\
  \bibinfo {pages} {045207} (\bibinfo {year} {2018})}\BibitemShut {NoStop}%
\bibitem [{\citenamefont {Schindler}\ \emph {et~al.}(2017)\citenamefont
  {Schindler}, \citenamefont {Regnault},\ and\ \citenamefont
  {Neupert}}]{PhysRevB.95.245134}%
  \BibitemOpen
  \bibfield  {author} {\bibinfo {author} {\bibfnamefont {F.}~\bibnamefont
  {Schindler}}, \bibinfo {author} {\bibfnamefont {N.}~\bibnamefont {Regnault}},
  \ and\ \bibinfo {author} {\bibfnamefont {T.}~\bibnamefont {Neupert}},\ }\href
  {\doibase 10.1103/PhysRevB.95.245134} {\bibfield  {journal} {\bibinfo
  {journal} {Phys. Rev. B}\ }\textbf {\bibinfo {volume} {95}},\ \bibinfo
  {pages} {245134} (\bibinfo {year} {2017})}\BibitemShut {NoStop}%
\bibitem [{\citenamefont {Zhang}\ \emph {et~al.}(2017)\citenamefont {Zhang},
  \citenamefont {Melko},\ and\ \citenamefont {Kim}}]{PhysRevB.96.245119}%
  \BibitemOpen
  \bibfield  {author} {\bibinfo {author} {\bibfnamefont {Y.}~\bibnamefont
  {Zhang}}, \bibinfo {author} {\bibfnamefont {R.~G.}\ \bibnamefont {Melko}}, \
  and\ \bibinfo {author} {\bibfnamefont {E.-A.}\ \bibnamefont {Kim}},\ }\href
  {\doibase 10.1103/PhysRevB.96.245119} {\bibfield  {journal} {\bibinfo
  {journal} {Phys. Rev. B}\ }\textbf {\bibinfo {volume} {96}},\ \bibinfo
  {pages} {245119} (\bibinfo {year} {2017})}\BibitemShut {NoStop}%
\bibitem [{\citenamefont {Zhang}\ and\ \citenamefont
  {Kim}(2017)}]{PhysRevLett.118.216401}%
  \BibitemOpen
  \bibfield  {author} {\bibinfo {author} {\bibfnamefont {Y.}~\bibnamefont
  {Zhang}}\ and\ \bibinfo {author} {\bibfnamefont {E.-A.}\ \bibnamefont
  {Kim}},\ }\href {\doibase 10.1103/PhysRevLett.118.216401} {\bibfield
  {journal} {\bibinfo  {journal} {Phys. Rev. Lett.}\ }\textbf {\bibinfo
  {volume} {118}},\ \bibinfo {pages} {216401} (\bibinfo {year}
  {2017})}\BibitemShut {NoStop}%
\bibitem [{\citenamefont {Broecker}\ \emph
  {et~al.}(2017{\natexlab{b}})\citenamefont {Broecker}, \citenamefont
  {Assaad},\ and\ \citenamefont {Trebst}}]{broecker2017quantum}%
  \BibitemOpen
  \bibfield  {author} {\bibinfo {author} {\bibfnamefont {P.}~\bibnamefont
  {Broecker}}, \bibinfo {author} {\bibfnamefont {F.~F.}\ \bibnamefont
  {Assaad}}, \ and\ \bibinfo {author} {\bibfnamefont {S.}~\bibnamefont
  {Trebst}},\ }\href@noop {} {\bibfield  {journal} {\bibinfo  {journal} {arXiv
  preprint arXiv:1707.00663}\ } (\bibinfo {year}
  {2017}{\natexlab{b}})}\BibitemShut {NoStop}%
\bibitem [{\citenamefont {Vargas-Hernández}\ \emph {et~al.}(2018)\citenamefont
  {Vargas-Hernández}, \citenamefont {Sous}, \citenamefont {Berciu},\ and\
  \citenamefont {Krems}}]{Rodrigo2018Extrapolating}%
  \BibitemOpen
  \bibfield  {author} {\bibinfo {author} {\bibfnamefont {R.~A.}\ \bibnamefont
  {Vargas-Hernández}}, \bibinfo {author} {\bibfnamefont {J.}~\bibnamefont
  {Sous}}, \bibinfo {author} {\bibfnamefont {M.}~\bibnamefont {Berciu}}, \ and\
  \bibinfo {author} {\bibfnamefont {R.~V.}\ \bibnamefont {Krems}},\ }\href@noop
  {} {\bibfield  {journal} {\bibinfo  {journal} {arXiv preprint
  arXiv:1803.08195}\ } (\bibinfo {year} {2018})}\BibitemShut {NoStop}%
\bibitem [{\citenamefont {Tanaka}\ and\ \citenamefont
  {Tomiya}(2017)}]{tanaka2017detection}%
  \BibitemOpen
  \bibfield  {author} {\bibinfo {author} {\bibfnamefont {A.}~\bibnamefont
  {Tanaka}}\ and\ \bibinfo {author} {\bibfnamefont {A.}~\bibnamefont
  {Tomiya}},\ }\href@noop {} {\bibfield  {journal} {\bibinfo  {journal}
  {Journal of the Physical Society of Japan}\ }\textbf {\bibinfo {volume}
  {86}},\ \bibinfo {pages} {063001} (\bibinfo {year} {2017})}\BibitemShut
  {NoStop}%
\bibitem [{\citenamefont {Landau}\ and\ \citenamefont
  {Binder}(2014)}]{mc_binder}%
  \BibitemOpen
  \bibfield  {author} {\bibinfo {author} {\bibfnamefont {D.~P.}\ \bibnamefont
  {Landau}}\ and\ \bibinfo {author} {\bibfnamefont {K.}~\bibnamefont
  {Binder}},\ }\href@noop {} {\emph {\bibinfo {title} {A Guide to Monte Carlo
  Simulations in Statistical Physics(4th Edition)}}}\ (\bibinfo  {publisher}
  {Cambridge University Press},\ \bibinfo {year} {2014})\BibitemShut {NoStop}%
\bibitem [{\citenamefont {Pollet}(2012)}]{qmc_review}%
  \BibitemOpen
  \bibfield  {author} {\bibinfo {author} {\bibfnamefont {L.}~\bibnamefont
  {Pollet}},\ }\href {http://stacks.iop.org/0034-4885/75/i=9/a=094501}
  {\bibfield  {journal} {\bibinfo  {journal} {Reports on Progress in Physics}\
  }\textbf {\bibinfo {volume} {75}},\ \bibinfo {pages} {094501} (\bibinfo
  {year} {2012})}\BibitemShut {NoStop}%
\bibitem [{not()}]{note}%
  \BibitemOpen
  \href@noop {} {}\bibinfo {note} {For Bose Hubbard model with $\beta = 100$,
  $t=0.03$, $U=1$, $L_x\times L_y=144$, and $N=64$, if the 20,000 samples are
  used, the total data size is about 700MB. However, the original configuration
  before compression has about 50,000 slides, the data size will be around
  500GB.}\BibitemShut {Stop}%
\bibitem [{\citenamefont {Sandvik}(1999)}]{sse1}%
  \BibitemOpen
  \bibfield  {author} {\bibinfo {author} {\bibfnamefont {A.~W.}\ \bibnamefont
  {Sandvik}},\ }\href {\doibase 10.1103/PhysRevB.59.R14157} {\bibfield
  {journal} {\bibinfo  {journal} {Phys. Rev. B}\ }\textbf {\bibinfo {volume}
  {59}},\ \bibinfo {pages} {R14157} (\bibinfo {year} {1999})}\BibitemShut
  {NoStop}%
\bibitem [{\citenamefont {Sylju\aa{}sen}\ and\ \citenamefont
  {Sandvik}(2002)}]{sse2}%
  \BibitemOpen
  \bibfield  {author} {\bibinfo {author} {\bibfnamefont {O.~F.}\ \bibnamefont
  {Sylju\aa{}sen}}\ and\ \bibinfo {author} {\bibfnamefont {A.~W.}\ \bibnamefont
  {Sandvik}},\ }\href {\doibase 10.1103/PhysRevE.66.046701} {\bibfield
  {journal} {\bibinfo  {journal} {Phys. Rev. E}\ }\textbf {\bibinfo {volume}
  {66}},\ \bibinfo {pages} {046701} (\bibinfo {year} {2002})}\BibitemShut
  {NoStop}%
\bibitem [{\citenamefont {Louis}\ and\ \citenamefont {Gros}(2004)}]{sse3}%
  \BibitemOpen
  \bibfield  {author} {\bibinfo {author} {\bibfnamefont {K.}~\bibnamefont
  {Louis}}\ and\ \bibinfo {author} {\bibfnamefont {C.}~\bibnamefont {Gros}},\
  }\href {\doibase 10.1103/PhysRevB.70.100410} {\bibfield  {journal} {\bibinfo
  {journal} {Phys. Rev. B}\ }\textbf {\bibinfo {volume} {70}},\ \bibinfo
  {pages} {100410} (\bibinfo {year} {2004})}\BibitemShut {NoStop}%
\bibitem [{\citenamefont {Sengupta}\ \emph {et~al.}(2002)\citenamefont
  {Sengupta}, \citenamefont {Sandvik},\ and\ \citenamefont {Campbell}}]{sse4}%
  \BibitemOpen
  \bibfield  {author} {\bibinfo {author} {\bibfnamefont {P.}~\bibnamefont
  {Sengupta}}, \bibinfo {author} {\bibfnamefont {A.~W.}\ \bibnamefont
  {Sandvik}}, \ and\ \bibinfo {author} {\bibfnamefont {D.~K.}\ \bibnamefont
  {Campbell}},\ }\href {\doibase 10.1103/PhysRevB.65.155113} {\bibfield
  {journal} {\bibinfo  {journal} {Phys. Rev. B}\ }\textbf {\bibinfo {volume}
  {65}},\ \bibinfo {pages} {155113} (\bibinfo {year} {2002})}\BibitemShut
  {NoStop}%
\bibitem [{\citenamefont {Becker}\ \emph {et~al.}(2010)\citenamefont {Becker},
  \citenamefont {Soltan-Panahi}, \citenamefont {Kronjger}, \citenamefont
  {Drscher}, \citenamefont {Bongs},\ and\ \citenamefont
  {Sengstock}}]{sengstock}%
  \BibitemOpen
  \bibfield  {author} {\bibinfo {author} {\bibfnamefont {C.}~\bibnamefont
  {Becker}}, \bibinfo {author} {\bibfnamefont {P.}~\bibnamefont
  {Soltan-Panahi}}, \bibinfo {author} {\bibfnamefont {J.}~\bibnamefont
  {Kronjger}}, \bibinfo {author} {\bibfnamefont {S.}~\bibnamefont {Drscher}},
  \bibinfo {author} {\bibfnamefont {K.}~\bibnamefont {Bongs}}, \ and\ \bibinfo
  {author} {\bibfnamefont {K.}~\bibnamefont {Sengstock}},\ }\href
  {http://stacks.iop.org/1367-2630/12/i=6/a=065025} {\bibfield  {journal}
  {\bibinfo  {journal} {New Journal of Physics}\ }\textbf {\bibinfo {volume}
  {12}},\ \bibinfo {pages} {065025} (\bibinfo {year} {2010})}\BibitemShut
  {NoStop}%
\bibitem [{\citenamefont {Fisher}\ \emph {et~al.}(1989)\citenamefont {Fisher},
  \citenamefont {Weichman}, \citenamefont {Grinstein},\ and\ \citenamefont
  {Fisher}}]{fisher}%
  \BibitemOpen
  \bibfield  {author} {\bibinfo {author} {\bibfnamefont {M.~P.~A.}\
  \bibnamefont {Fisher}}, \bibinfo {author} {\bibfnamefont {P.~B.}\
  \bibnamefont {Weichman}}, \bibinfo {author} {\bibfnamefont {G.}~\bibnamefont
  {Grinstein}}, \ and\ \bibinfo {author} {\bibfnamefont {D.~S.}\ \bibnamefont
  {Fisher}},\ }\href {\doibase 10.1103/PhysRevB.40.546} {\bibfield  {journal}
  {\bibinfo  {journal} {Phys. Rev. B}\ }\textbf {\bibinfo {volume} {40}},\
  \bibinfo {pages} {546} (\bibinfo {year} {1989})}\BibitemShut {NoStop}%
\bibitem [{\citenamefont {Hu}\ \emph {et~al.}()\citenamefont {Hu},
  \citenamefont {Santos}, \citenamefont {Pelster}, \citenamefont {Eggert},\
  and\ \citenamefont {Zhang}}]{AF_tri}%
  \BibitemOpen
  \bibfield  {author} {\bibinfo {author} {\bibfnamefont {S.-J.}\ \bibnamefont
  {Hu}}, \bibinfo {author} {\bibfnamefont {F.~E. A.~d.}\ \bibnamefont
  {Santos}}, \bibinfo {author} {\bibfnamefont {A.}~\bibnamefont {Pelster}},
  \bibinfo {author} {\bibfnamefont {S.}~\bibnamefont {Eggert}}, \ and\ \bibinfo
  {author} {\bibfnamefont {X.-F.}\ \bibnamefont {Zhang}},\ }\href@noop {}
  {\bibinfo  {journal} {in preparation}\ }\BibitemShut {NoStop}%
\bibitem [{\citenamefont {Zhang}\ \emph {et~al.}(2011)\citenamefont {Zhang},
  \citenamefont {Dillenschneider}, \citenamefont {Yu},\ and\ \citenamefont
  {Eggert}}]{ss10}%
  \BibitemOpen
\bibfield  {journal} {  }\bibfield  {author} {\bibinfo {author} {\bibfnamefont
  {X.-F.}\ \bibnamefont {Zhang}}, \bibinfo {author} {\bibfnamefont
  {R.}~\bibnamefont {Dillenschneider}}, \bibinfo {author} {\bibfnamefont
  {Y.}~\bibnamefont {Yu}}, \ and\ \bibinfo {author} {\bibfnamefont
  {S.}~\bibnamefont {Eggert}},\ }\href {\doibase 10.1103/PhysRevB.84.174515}
  {\bibfield  {journal} {\bibinfo  {journal} {Phys. Rev. B}\ }\textbf {\bibinfo
  {volume} {84}},\ \bibinfo {pages} {174515} (\bibinfo {year}
  {2011})}\BibitemShut {NoStop}%
\bibitem [{\citenamefont {Wessel}\ and\ \citenamefont {Troyer}(2005)}]{ss1}%
  \BibitemOpen
  \bibfield  {author} {\bibinfo {author} {\bibfnamefont {S.}~\bibnamefont
  {Wessel}}\ and\ \bibinfo {author} {\bibfnamefont {M.}~\bibnamefont
  {Troyer}},\ }\href {\doibase 10.1103/PhysRevLett.95.127205} {\bibfield
  {journal} {\bibinfo  {journal} {Phys. Rev. Lett.}\ }\textbf {\bibinfo
  {volume} {95}},\ \bibinfo {pages} {127205} (\bibinfo {year}
  {2005})}\BibitemShut {NoStop}%
\bibitem [{\citenamefont {Heidarian}\ and\ \citenamefont {Damle}(2005)}]{ss2}%
  \BibitemOpen
  \bibfield  {author} {\bibinfo {author} {\bibfnamefont {D.}~\bibnamefont
  {Heidarian}}\ and\ \bibinfo {author} {\bibfnamefont {K.}~\bibnamefont
  {Damle}},\ }\href {\doibase 10.1103/PhysRevLett.95.127206} {\bibfield
  {journal} {\bibinfo  {journal} {Phys. Rev. Lett.}\ }\textbf {\bibinfo
  {volume} {95}},\ \bibinfo {pages} {127206} (\bibinfo {year}
  {2005})}\BibitemShut {NoStop}%
\bibitem [{\citenamefont {Melko}\ \emph {et~al.}(2005)\citenamefont {Melko},
  \citenamefont {Paramekanti}, \citenamefont {Burkov}, \citenamefont
  {Vishwanath}, \citenamefont {Sheng},\ and\ \citenamefont {Balents}}]{ss3}%
  \BibitemOpen
  \bibfield  {author} {\bibinfo {author} {\bibfnamefont {R.~G.}\ \bibnamefont
  {Melko}}, \bibinfo {author} {\bibfnamefont {A.}~\bibnamefont {Paramekanti}},
  \bibinfo {author} {\bibfnamefont {A.~A.}\ \bibnamefont {Burkov}}, \bibinfo
  {author} {\bibfnamefont {A.}~\bibnamefont {Vishwanath}}, \bibinfo {author}
  {\bibfnamefont {D.~N.}\ \bibnamefont {Sheng}}, \ and\ \bibinfo {author}
  {\bibfnamefont {L.}~\bibnamefont {Balents}},\ }\href {\doibase
  10.1103/PhysRevLett.95.127207} {\bibfield  {journal} {\bibinfo  {journal}
  {Phys. Rev. Lett.}\ }\textbf {\bibinfo {volume} {95}},\ \bibinfo {pages}
  {127207} (\bibinfo {year} {2005})}\BibitemShut {NoStop}%
\bibitem [{\citenamefont {Boninsegni}\ and\ \citenamefont
  {Prokof'ev}(2005)}]{ss4}%
  \BibitemOpen
  \bibfield  {author} {\bibinfo {author} {\bibfnamefont {M.}~\bibnamefont
  {Boninsegni}}\ and\ \bibinfo {author} {\bibfnamefont {N.}~\bibnamefont
  {Prokof'ev}},\ }\href {\doibase 10.1103/PhysRevLett.95.237204} {\bibfield
  {journal} {\bibinfo  {journal} {Phys. Rev. Lett.}\ }\textbf {\bibinfo
  {volume} {95}},\ \bibinfo {pages} {237204} (\bibinfo {year}
  {2005})}\BibitemShut {NoStop}%
\bibitem [{\citenamefont {Sen}\ \emph {et~al.}(2008)\citenamefont {Sen},
  \citenamefont {Dutt}, \citenamefont {Damle},\ and\ \citenamefont
  {Moessner}}]{ss5}%
  \BibitemOpen
  \bibfield  {author} {\bibinfo {author} {\bibfnamefont {A.}~\bibnamefont
  {Sen}}, \bibinfo {author} {\bibfnamefont {P.}~\bibnamefont {Dutt}}, \bibinfo
  {author} {\bibfnamefont {K.}~\bibnamefont {Damle}}, \ and\ \bibinfo {author}
  {\bibfnamefont {R.}~\bibnamefont {Moessner}},\ }\href {\doibase
  10.1103/PhysRevLett.100.147204} {\bibfield  {journal} {\bibinfo  {journal}
  {Phys. Rev. Lett.}\ }\textbf {\bibinfo {volume} {100}},\ \bibinfo {pages}
  {147204} (\bibinfo {year} {2008})}\BibitemShut {NoStop}%
\bibitem [{\citenamefont {Wang}\ \emph {et~al.}(2009)\citenamefont {Wang},
  \citenamefont {Pollmann},\ and\ \citenamefont {Vishwanath}}]{ss6}%
  \BibitemOpen
  \bibfield  {author} {\bibinfo {author} {\bibfnamefont {F.}~\bibnamefont
  {Wang}}, \bibinfo {author} {\bibfnamefont {F.}~\bibnamefont {Pollmann}}, \
  and\ \bibinfo {author} {\bibfnamefont {A.}~\bibnamefont {Vishwanath}},\
  }\href {\doibase 10.1103/PhysRevLett.102.017203} {\bibfield  {journal}
  {\bibinfo  {journal} {Phys. Rev. Lett.}\ }\textbf {\bibinfo {volume} {102}},\
  \bibinfo {pages} {017203} (\bibinfo {year} {2009})}\BibitemShut {NoStop}%
\bibitem [{\citenamefont {Heidarian}\ and\ \citenamefont
  {Paramekanti}(2010)}]{ss7}%
  \BibitemOpen
  \bibfield  {author} {\bibinfo {author} {\bibfnamefont {D.}~\bibnamefont
  {Heidarian}}\ and\ \bibinfo {author} {\bibfnamefont {A.}~\bibnamefont
  {Paramekanti}},\ }\href {\doibase 10.1103/PhysRevLett.104.015301} {\bibfield
  {journal} {\bibinfo  {journal} {Phys. Rev. Lett.}\ }\textbf {\bibinfo
  {volume} {104}},\ \bibinfo {pages} {015301} (\bibinfo {year}
  {2010})}\BibitemShut {NoStop}%
\bibitem [{\citenamefont {Bonnes}\ and\ \citenamefont {Wessel}(2011)}]{ss8}%
  \BibitemOpen
  \bibfield  {author} {\bibinfo {author} {\bibfnamefont {L.}~\bibnamefont
  {Bonnes}}\ and\ \bibinfo {author} {\bibfnamefont {S.}~\bibnamefont
  {Wessel}},\ }\href {\doibase 10.1103/PhysRevB.84.054510} {\bibfield
  {journal} {\bibinfo  {journal} {Phys. Rev. B}\ }\textbf {\bibinfo {volume}
  {84}},\ \bibinfo {pages} {054510} (\bibinfo {year} {2011})}\BibitemShut
  {NoStop}%
\bibitem [{\citenamefont {Yamamoto}\ \emph {et~al.}(2013)\citenamefont
  {Yamamoto}, \citenamefont {Ozaki}, \citenamefont {S\'a~de Melo},\ and\
  \citenamefont {Danshita}}]{ss9}%
  \BibitemOpen
  \bibfield  {author} {\bibinfo {author} {\bibfnamefont {D.}~\bibnamefont
  {Yamamoto}}, \bibinfo {author} {\bibfnamefont {T.}~\bibnamefont {Ozaki}},
  \bibinfo {author} {\bibfnamefont {C.~A.~R.}\ \bibnamefont {S\'a~de Melo}}, \
  and\ \bibinfo {author} {\bibfnamefont {I.}~\bibnamefont {Danshita}},\ }\href
  {\doibase 10.1103/PhysRevA.88.033624} {\bibfield  {journal} {\bibinfo
  {journal} {Phys. Rev. A}\ }\textbf {\bibinfo {volume} {88}},\ \bibinfo
  {pages} {033624} (\bibinfo {year} {2013})}\BibitemShut {NoStop}%
\bibitem [{\citenamefont {Pollock}\ and\ \citenamefont
  {Ceperley}(1987)}]{rhos}%
  \BibitemOpen
  \bibfield  {author} {\bibinfo {author} {\bibfnamefont {E.~L.}\ \bibnamefont
  {Pollock}}\ and\ \bibinfo {author} {\bibfnamefont {D.~M.}\ \bibnamefont
  {Ceperley}},\ }\href {\doibase 10.1103/PhysRevB.36.8343} {\bibfield
  {journal} {\bibinfo  {journal} {Phys. Rev. B}\ }\textbf {\bibinfo {volume}
  {36}},\ \bibinfo {pages} {8343} (\bibinfo {year} {1987})}\BibitemShut
  {NoStop}%
\bibitem [{\citenamefont {Hsu}\ \emph {et~al.}(2018)\citenamefont {Hsu},
  \citenamefont {Li}, \citenamefont {Deng},\ and\ \citenamefont
  {Das~Sarma}}]{hsu2018machine}%
  \BibitemOpen
  \bibfield  {author} {\bibinfo {author} {\bibfnamefont {Y.-T.}\ \bibnamefont
  {Hsu}}, \bibinfo {author} {\bibfnamefont {X.}~\bibnamefont {Li}}, \bibinfo
  {author} {\bibfnamefont {D.-L.}\ \bibnamefont {Deng}}, \ and\ \bibinfo
  {author} {\bibfnamefont {S.}~\bibnamefont {Das~Sarma}},\ }\href@noop {}
  {\bibfield  {journal} {\bibinfo  {journal} {arXiv preprint arXiv:1805.12138}\
  } (\bibinfo {year} {2018})}\BibitemShut {NoStop}%
\end{thebibliography}%

\end{document}